\documentclass[useAMS,usenatbib]{mn2e}
\usepackage{wasysym}
\usepackage{epsfig}
\usepackage{multirow}
\usepackage{color}

\usepackage{amssymb,amsmath}
\usepackage{graphicx}
\usepackage{natbib}

\definecolor{gold}{rgb}{1.,0.647059,0.}
 \definecolor{purple}{rgb}{0.627451,0.12549,0.941176}
 \definecolor{darkgreen}{rgb}{0.133333,0.545098,0.133333}
\definecolor{marroncaca}{rgb}{0.6,0.2,0}
\definecolor{orangee}{rgb}{1.,0.6,0.4}
\definecolor{olivegreen}{rgb}{0.419608,0.556863,0.137255}

\newcommand{\aap}{A\&A}

\title{Variability of X-ray binaries from an oscillating {hot} corona.}
\author[C. Cabanac, G. Henri, P.-O. Petrucci, J. Malzac, J. Ferreira, T. M. Belloni]{C. Cabanac$^{1,2,3}$\thanks{E-mail:
cabanac@cesr.fr (CC)}, G.Henri$^{3}$, P.-~O Petrucci$^{3}$, J. Malzac$^{1}$, J. Ferreira$^{3}$ and T. M. Belloni$^{4}$\\
 $^{1}$Centre d'\'Etude Spatiale des Rayonnements, CNRS-UPS, 9 Avenue du Colonel Roche, 31028 Toulouse Cedex 4, France\\
$^{2}$School of Physics and Astronomy, University of Southampton, Southampton SO17 1BJ, UK\\ $^{3}$Laboratoire d'Astrophysique de Grenoble--Universit\'e Joseph-Fourier/CNRS UMR 5571 --BP~53, F-38041 Grenoble, France \\ $^{4}$INAF-Osservatorio Astronomico di Brera, Via E. Bianchi 46, I-23807 Merate (LC), Italy
}

\begin{document}
\date{Accepted 2010 January 12. Received 2010 January 12; in original form 2009 September 09}

\pagerange{\pageref{firstpage} -- \pageref{LastPage}} \pubyear{2009}

\maketitle

\label{firstpage}

\begin{abstract}
The spectral and timing properties of an oscillating hot thermal corona are investigated. This oscillation is assumed to be due to a magneto-acoustic wave propagating within the corona and triggered by an external, non specified, excitation. A cylindrical geometry is adopted and, neglecting the rotation, the wave equation is solved in for different boundary conditions.   The resulting X-ray luminosity,  through thermal comptonization of embedded soft photons, is then computed, first analytically, assuming linear dependence between the local pressure disturbance and the radiative modulation. These calculations are also compared to  Monte-Carlo simulations. The main results of this study are: (1) the corona plays the role of a low band-pass medium, its response to a white noise excitation being  a  flat top noise Power Spectral Density (PSD) at low frequencies and  a red noise at high frequency, (2) resonant peaks are present in the PSD. Their powers depend on the boundary conditions chosen and, more specifically, on the impedance adaptation with the external medium at the corona inner boundary. (3) The flat top noise level and break as well as the resonant peak frequencies are inversely proportional to the external radius $r_{\rm j}$. (4) Computed rms and f-spectra exibit an overall increase of the variability with energy. Comparison with observed variability features, especially in the hard intermediate states of X-ray binaries are discussed. \end{abstract}


\begin{figure*}
\includegraphics[width=1.\textwidth]{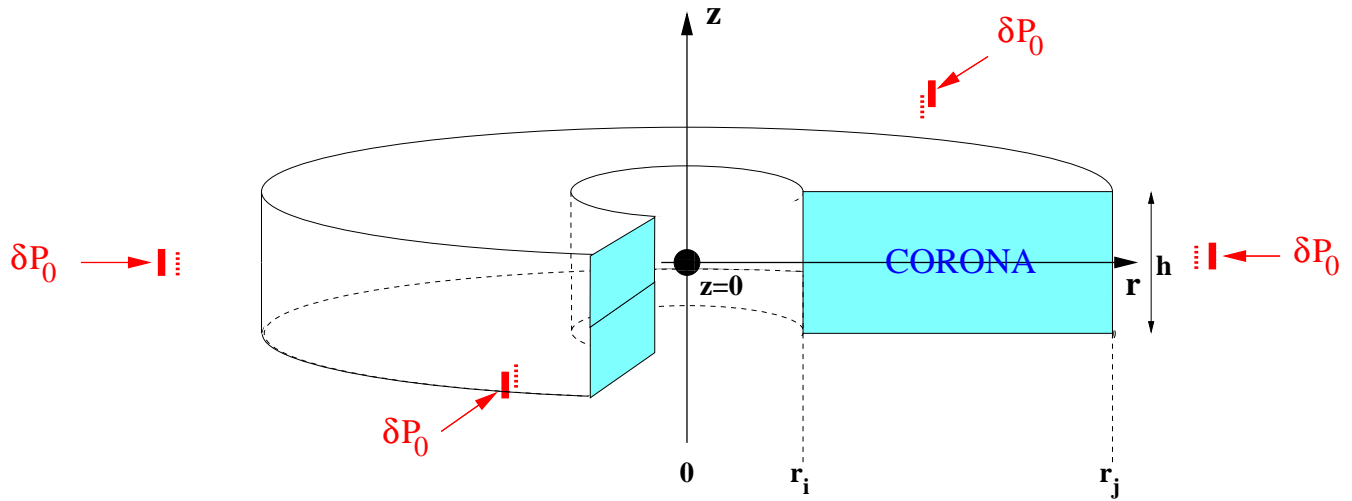}
\caption{Sketch of the model geometry. The corona is assumed to harbour cylindrical symmetry. It extends from its inner radius $r_{\rm i}$ to  its outer radius $r_{\rm j}$ and its height is $h_{\rm c}$.  Non-specified white noise excitations are assumed to occur at $r_{\rm j}$ then triggering (magneto-)sonic waves in the corona.  The resulting X-ray luminosity is then computed through thermal comptonization of embedded soft photons.}
\label{fig:sketch-model}
\end{figure*}
\section{Introduction}
X-ray binaries (XRBs) exhibit large variability on various timescales. While their spectral states and their accretion rates are typically changing from weeks to days (see e.g. the different canonical states observed in black hole binaries, as defined by \citealp{mcclintock2003}, \citealp{homan2005}, \citealp{belloni2009} and references therein), their light curves exhibit drastic changes from hours to millisecond. Several tools are now used in order to analyse these timing features (see e.g. \citealp{vklis2004} for a review) such as e.g. cross-correlation between different energy band (which allows to infer the so-called \emph{time-lags}), but the most commonly used is still the Fourier analysis via the computation of Power Spectral Distribution (hereafter PSD).\\
\indent Despite its known limitations (e.g. signal phase lost in the analysis), any attempt to model the physics of XRBs has to take into account the various features observed in the PSD and its evolution when the source transits from a state to another. In the so-called hard state, the level of the variability is high and the PSD harbour a Band Limited Noises (BLNs) shape extending up to a break frequency $\nu_b$. Large peaks, the so called Quasi-Periodic Oscillations (QPO), can also be observed. In the soft state the Poissonian noise is usually dominating on all the frequency range and no or weak QPOs are detected. Different types of QPOs (called A, B and C) can be identified depending on the value of their frequency, their strength and even their time lags (see again \citealp{vklis2004} or \citealp{casella2004} and references therein). However, if QPOs are remarkable features, the major part of the variability is usually aperiodic. As for energy spectra, variability evolves during time: in general, the overall variability decreases as the spectrum softens, with frequencies increasing, until in the soft state where the BLN reaches a low level (see e.g. \citealp{belloni2005, belloni2009}).\\

\indent Several models intending to interpret variability features focuses on the QPO phenomena. For the high frequency QPOs ($\nu>100\ \rm Hz$), lense-thirring precession (see e.g. \citealp{stella1998}) or beating frequency between particular orbits \citep{lamb2003} have been proposed. For the low frequency QPOs ($0.01 <\nu<100\ \rm Hz$), fewer models are available (see e.g. \citealp{tagger1999} or \citealp{titarchuk2005}, but also \citealp{stella1998}), but they usually try to explain the observed correlation between frequencies and/or other observables, without taking into account the whole emitting process. However, it is worth noting that it is in the highest energy bands that the X-ray flux is observed as being highly variable. In contrast in most of the available models, the  source of variability lies in the geometrically thin accretion disc, which emit mainly at lower energies. Note that some observations also suggest directly that the disc is less variable than the corona (see e.g. \citealp{churazov01}, \citealp{rodriguez2003} however see also \citealp{wilkinson09}). In consequence a proper model for variability in BHB has to deal with the radiative transfer between the disc and the corona.  \\

\indent A few models have already been proposed in the literature. For instance, a full modelling of the BLN component has been attempted by \cite{misra2000} in the framework of a ``transition disc model'': following \cite{nowak1999}'s idea, an acoustic wave is propagating within the accretion disc. In this framework, the author considers only the propagation to occur in one direction towards a central sink. This model is then used to explain qualitatively the general shape of the power spectra and lag energy dependencies observed in Cyg X-1. \\
 \cite{psaltis2000} tried to model the filter effect of a narrow annulus in a geometrically thin disc. However, the nature of this annulus is not specified and its typical extension is $\delta r/r<10^{-2}$. They also neglect the contribution of the radial pressure forces. In this framework and depending on the mode of oscillation chosen, the external perturbation is shown to be modulated in amplitude according to the exciting frequency.  The square of the response in pressure exhibits Lorentzians which could account for the observed behaviour in BH or NS binaries. In a similar spirit, \cite{lee2004} studied numerically the response to radial perturbations in an accretion torus. They showed that resonances could occur and become larger when the frequency difference between the radial and vertical epicyclic frequencies where half the forcing frequencies. 
\\  
\indent One of the very first attempt to evaluate the effect of propagation in a comptonising region on the timing behaviour was examined by \cite{miyamoto1988}. They showed that the tight period dependence of the observed time lags in Cyg X-1 could not be accounted for by the inverse Compton scatterings process only. In another framework, \cite{zycki2005} tested their model of variability \citep{zycki2003} where the variable emission responsible for the noise component is attributed to multiple active regions/perturbations moving radially towards the central black hole. The QPOs are obtained by modulating either the reflection amplitude, the heating rate, the covering factor of the reprocessor or the column density value. They predict for each of these cases the corresponding power and f spectra (for a definition of f-spectra, see \citealp{revnivtsev1999}), and the time lag energy dependencies. In a subsequent paper, \cite{sobolewska2006} tested their model on real data and concluded that within this framework the QPO spectra in the hard state are always softer than normal average spectra and cold disc oscillations might then be responsible for the low frequency QPO. On the other hand, when the energey spectra gets softer, the QPO spectra are harder and the low frequency QPO might then originate from the hot plasma.\\
\indent In a more recent study, {\cite{schnittman2006} try to model the oscillation of a torus in Kerr metric, via three radiative processes: a thin emission line then a thick one, and finally an optically thick thermal emission process. They manage to reproduce some of the the properties of C-type QPOs, especially the observed increase of the amplitude with the inclination of the system.}\\
\indent In this paper we present a new approach that deals with the radiative transfer into an oscillating corona. These oscillations are assumed to be due to a magneto-acoustic wave propagating within the corona, modulating the efficiency of the comptonisation process on embedded soft photons. These basic ingredients give a promising framework to reproduce the main timing features of the X-ray binary in hard and hard-intermediate states such as the Band Limited Noise continuum and C-type LFQPO. The assumptions of the models and wave equation solutions are detailed in section \ref{sec:model}. An analytical study in the linear approximation is discussed in section \ref{sec:linear} and compared with Monte-Carlo simulations in section \ref{sec:monte-carlo}. We finally discuss the main results of this toy model and the comparison to observation in section \ref{sec:discussion}.
\section{The model}\label{sec:model}
\subsection{Basic assumptions}
The general structure of the model is sketched in Fig. \ref{fig:sketch-model}. A hot optically thin cylindrical medium is assumed, hereafter called the ``corona''. It is limited radially by its inner  and outer radii  $r_{\rm i}$ and $r_{\rm j}$ respectively. This corona has thus a ring shape of typical height $h_{\rm c}$. This geometry agrees with the observations that suggest the X-ray corona to be closely linked to the base of the jet in XRB (e.g. \citealt{markoff2001, fender2004, markoff2005}). Then the inner radius $r_{\rm i}$ can be identified as the last stable orbit close to the central compact object. On the other hand $r_{\rm j}$ can be compared to the transition radius between an outer standard accretion disc and the inner hot corona, a geometry commonly invoked in XRB (e.g. \citealp{esin1997}, \citealp{ferreira2006}, \citealp{done2007}). \\
\indent In this paper, no hypothesis is done on the physical origin of these radii. For simplicity, the corona is assumed to have, at rest, a constant temperature $T_0$ and density $n_0$ and consequently a constant pressure $P_0$. We then consider that pressure instabilities at the external radius $r_{\rm j}$ generate a sound wave within this thermalised plasma. The origin and the nature of those instabilities are not discussed in this paper since we focus only on the radiative response of the corona. We assume these instabilities to have a white noise spectrum (i.e. same amplitude for all excitation frequencies) which corresponds to a Dirac perturbation in the temporal domain, i.e, we limit our study to the corona transfert function . For seek of simplicity we also restrict our calculation to the 1D case i.e. the wave will only propagate radially at the sound velocity:
\begin{equation}
c_{\rm s} = \sqrt{\frac{kT_0}{m_p}}\simeq 3.1\times 10^8 \left (\frac{T_0}{100\ \mbox{keV}}\right )^{1/2} \mbox{cm.s$^{-1}$}.\label{eq:cs-kt}
\end{equation}
In the last expression $T_0$ is in keV. It is important to note that there are simplifications which make this paper only a first step. Once additional complications such as incorporating the effects of rotation are introduced, they might lead to some revision of the scenario presented here. This will be addressed in a forthcoming paper. Note however that the nature of the corona and hence its exact rotation profile is still mainly unknown.\\
 Finally we suppose a blackbody seed photon field of temperature $T_{seed}$ emitted isotropically at the corona midplane (i.e. in $z=0$ see Fig. \ref{fig:sketch-model}). These seed photons will be comptonized in the corona then producing a variable X-ray emission.
\subsection{Wave equation}
In cylindrical 1D geometry, in absence of local damping or excitation, and neglecting the rotation, the basic wave propagation equation can be written as:
\begin{equation}
 \frac{1}{r}\frac{\partial}{\partial r}\left(r\frac{\partial p}{\partial
   r}\right)
 - \frac{1}{c_{\rm s}^2}\frac{\partial^2 p}{\partial t^2} = 0,\label{equ_onde2}.
\end{equation}
$p(r,t)$ is the perturbation in pressure given by, in complex notation and for a given frequency $\nu=\omega/2\pi$:
\begin{equation}
p(r,t) = p_r(r)\exp^{\imath(- \omega t)}.\label{solution}
\end{equation}
Introducing the new variable $\displaystyle x = \frac{\omega}{c_{\rm s}}r$ and putting Eq. \ref{solution} in Eq. \ref{equ_onde2}, we get:
\begin{equation}
x^2\frac{d^2\,p_r}{d\,x^2} + x\frac{d\,p_r}{d\,x} + x^2 p_r = 0 \label{equ_bessel1}
\end{equation}
The general solution of those equations are linear combination of Hankel's function of the zeroth order $H_0^1$ and $H_0^2$ \citep{abramowitz1964}:
\begin{figure*}
\includegraphics[width=1.\textwidth]{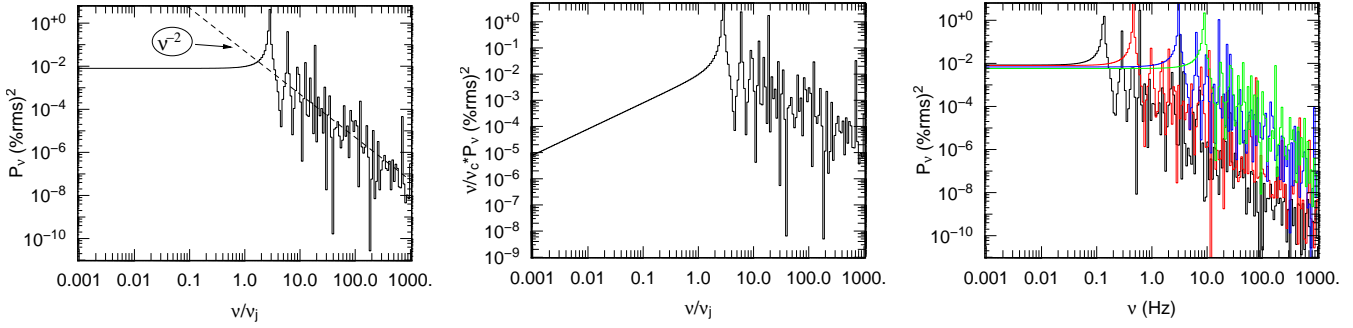}
\caption{{\it Left}: Power spectrum plotted in reduced frequency $x_j=\nu/\nu_j$ (with $\nu_j=2\pi c_s/r_j$) for a single zone corona with $\xi=10^{-2}$, $C_{lin}=1$ and $\delta P_0/P_0=0.1$. {\it Center}: same spectrum plotted in $\nu P_{\nu}$. {\it Right}: PSD plotted in true frequency and its evolution when moving $r_j$ inwards only. Here, $M=10\rm\ M_{\odot}$, $r_i= 2\ \rm R_g = 3\times 10^6 \ \rm cm$ and $kT_0=100\ \rm keV$ (hence $c_s=3.1\times 10^8\ \rm cm.s^{-1}$). Black: $r_{\rm j} = 680 \ \rm R_g = 10^9 \ \rm cm$, red: $r_{\rm j} = 200 \ \rm R_g = 3\times 10^8 \ \rm cm$,
 blue: $r_{\rm j} = 68 \ \rm R_g = 10^8 \ \rm cm$, green: $r_{\rm j} = 14 \ \rm R_g = 2\times 10^7 \ \rm cm$. (Note that $(\% rms)^2 (@\ \nu=0)$ slightly decreases and $\nu_{\rm b}$ increases when the external radius of the corona decreases, though the effect is tiny here.).}
\label{fig:PSD-anal}
\end{figure*}
\begin{equation}
p(x,t) = \left[\alpha H_0^1(x) + \beta
H_0^2(x)\right] e^{- \imath\omega t},\label{solu_J0} 
\end{equation}
$\alpha$ and $\beta$ being determined by the boundary conditions. 
Note that the above solution is similar to the one obtain by \cite{nowak1999}, or \cite{titarchuk2005} despite the fact that those authors used only one instead of a linear combination of both Hankel's function as solution.

\subsection{Boundary conditions: total reflection at the internal radius}\label{subsec:total-refl}
At the external radius of the corona $r_{\rm j}$, we assume a constant excitation $p_{0}$\footnote{Note the difference between $p_0$, the pressure perturbation imposed in $r=r_{\rm j}$ and $P_0$ the corona pressure at rest} (white noise hypothesis):
\begin{equation}
p(x_{\rm j},t) = \left(\alpha H_0^1(x_j) + \beta
H_0^2(x_j)\right) e^{- \imath \omega t} = p_0 e^{-\imath \omega t}.\label{eq:cond-lim-rad1}
\end{equation}
where $x_j=x(r=r_j)$. The closure relationship will be given by the behaviour of the wave at the internal radius $r_{\rm i}$. We will, as a primary assumption, consider that there is no transmission of the wave and hence total reflection in $r_{\rm i}$. If $r_{\rm i}$ is equal or close to the Last Stable Orbit, one would expect indeed the density of the corona to drop quickly inside this radius and hence the pressure as well. 
It thus gives:
\begin{equation}
p(x_{\rm i},t) = \left(\alpha H_0^1(x_i) + \beta
H_0^2(x_i)\right) e^{- \imath \omega t} = 0.\label{eq:cond-lim-rad2}
\end{equation}
where $x_i=x(r=r_i)$.The system of Eqs. \ref{eq:cond-lim-rad1} and \ref{eq:cond-lim-rad2} can be solved in order to obtain $\alpha$ and $\beta$:
\begin{eqnarray}
\alpha &=& p_0\frac{H_0^2(x_i)}{H_0^2(x_i)H_0^1(x_j)-H_0^1(x_i)H_0^2(x_j)}\label{eq:alpha}\\
\beta &=& p_0\frac{-H_0^1(x_i)}{H_0^2(x_i)H_0^1(x_j)-H_0^1(x_i)H_0^2(x_j)}\label{eq:beta}
\end{eqnarray}
\section{Analytical solutions in the linear and zero-phase approximations}\label{sec:linear}
In this section, we infer the shape of the expected PSD in the simple case where the radiative response of the corona depends linearly on the local perturbation. This appears to give analytical results in good agreement with our Monte Carlo simulations detailed in the next section.\\
\indent The corona being optically thin, the luminosity $dL_0$ emitted locally at rest in a ring of radius $r$, width $dr$ and height $h_c$  is proportional to its emissivity per unit volume $\eta_0$:
\begin{equation}
dL_0(r)=\eta_0 2\pi rdrh_c. \label{eq:emissiv-flux}
\end{equation}
In presence of the sonic wave, the luminosity $dL(r,t)$ of this ring varies in time and we will make the assumption that its relative variation $(dL-dL_0)/dL_0$ is a linear function of the relative pressure perturbation $p(r,t)/P_0$:
\begin{equation}
\frac{dL(r,t)-dL_0(r)}{dL_0(r)}=C_{lin}\frac{p(r,t)}{P_0}.\label{eq:linear-flux-pressure}
\end{equation}
$C_{lin}$ is the constant of proportionality which is in complete generality a function of the perturbation frequency.
However, if the delays implied by the multiple diffusions of the photon inside the corona are low compared to the period of the wave, little phase delay is expected between the pressure wave and the luminosity. Hence $C_{lin}$ will be a real number independent of $\omega$. This is what we call the ``zero-phase approximation''.  It is usually verified in optically thin plasma where photons travel a few $h_c$ before escaping (see e.g. \citealt{malzac2000}). Then a rough estimate of the time spent by the photons inside the corona compared to the period $2\pi/\omega$ of the wave  gives:
\begin{equation}
\frac{\omega}{2\pi}\frac{h_c}{c}=\frac{1}{2\pi}\frac{h_c}{r}\frac{c_s}{c}x
\end{equation}
It is generally much smaller than 1 in the cases we are interested in (i.e. corona aspect ratio $h_c/r<1$ and corona temperature of a few tens to hundred of keV) unless $x$ becomes of the order of 100 or 1000.\\

\noindent Combining Eqs  \ref{eq:emissiv-flux} and \ref{eq:linear-flux-pressure}, we obtain:
\begin{equation}
dL(r,t)=\eta_0\left(1+C_{lin}\frac{p(r,t)}{P_0}\right)2\pi r dr h_c.
\end{equation} 
Previous equation integrated on the whole volume of the corona gives:
\begin{equation}
 L=L_0+\underbrace{\eta_0\frac{2\pi h_c C_{lin}}{P_0}\int_{r_i}^{r_j}{p(r,t)rdr}}_{\equiv L_{\sim}},\label{eq:integ-flux-pressure}
\end{equation} 
with $\displaystyle L_0=\pi\eta_0 h_c (r_{\rm j}^2-r_{\rm i}^2)=\pi\eta_0 h_c x_j^2\frac{c_s^2}{\omega^2}(1-\xi^2)$, and: 
\begin{equation}
\xi\equiv\frac{r_i}{r_j}=\frac{x_i}{x_j}.
\end{equation}
Note that relationship \ref{eq:integ-flux-pressure} is only valid if the emissivity per unit volume $\eta_0$ of the corona, at rest, is uniform, which is a direct consequence of our assumptions of constant temperature and density $T_0$ and $n_0$. \\  
\subsection{Power Spectra obtained with large vertical wavelength and total reflection in $r_{\rm i}$} \label{sec:PSD-anal} 
Using Eq. \ref{solu_J0}, the expression of the perturbation in luminosity $L_{\sim}$ becomes
\begin{equation}
 L_{\sim}=\frac{2 L_0 C_{lin}}{P_0}\frac{e^{-i\omega t}}{x_j^2(1-\xi^2)}\int^{x_j}_{\xi x_j}\left(\alpha H_0^1\left(x\right)+\beta H_0^2\left(x\right)\right) xdx.\label{eq:integ-flux-pressure-longwave}
\end{equation}
The Hankel's functions present in $\alpha$ and $\beta$ (see Eqs. \ref{eq:alpha} and \ref{eq:beta}) are easily integrated (see \citealp{abramowitz1964}) and hence the previous equation leads to:
\begin{equation}
L_{\sim}=  \frac{2 L_0C_{lin}p_0}{P_0}\frac{e^{-i\omega t}}{x_j^2(1-\xi^2)} M_{x_{\rm j},\xi} \label{eq:calcul-analytique-composante-module-final1}
\end{equation}
with,
\begin{equation}
M_{x_j,\xi}=x_j \frac{H^2_{0,\rm i}(H^1_{1,\rm j}-\xi H^1_{1,\rm i})-H^1_{0,\rm i}(H^2_{1,\rm j}-\xi H^2_{1,\rm i})}{H^2_{0,\rm i}H^1_{0,\rm j}-H^1_{0,\rm i}H^2_{0,\rm j}}.\label{eq:calcul-analytique-composante-module-final2}
\end{equation}
$H_1^2$ and $H_1^1$ are the Hankel's function of first order and the index $\rm i$ or $\rm j$ corresponds to the point where the Hankel's function is evaluated i.e. $x_i$ or $x_j$ respectively. It is also easy to demonstrate that the function $M$ is real in case of total reflection (both numerator and denominator are pure imaginaries and hence the ratio is real). \\

\indent The PSD $\displaystyle P_{\nu}\equiv |L_{\sim}|^2/L_0^2$ can then be directly deduced from Eq. \ref{eq:calcul-analytique-composante-module-final1}:
\begin{equation}
P_{\nu}\equiv |L_{\sim}|^2/L_0^2= \left|{2 C_{lin}}\frac{p_0}{P_0}\frac{M(x_j,\xi)}{x_j^2(1-\xi^2)}\right|^2.\label{eq:psd-anal}
\end{equation}
It is directly proportional, due to our linear assumption, to the input perturbation through the term ${2C_{lin}p_0}/{P_0}$ but it is also modulated by the intrinsic response of the corona through the function ${M(x_j,\xi)}/({x_j^2(1-\xi^2)})$.

In the zero-phase approximation (see above) $C_{lin}$ is real and we can find asymptotic expressions of the PSD  for low and high frequencies. For low frequencies, the function $M$ can be approximated to (see appendix \ref{app:A}):
\begin{equation}
M(x_j,\xi) \underset{x_j \to 0} {\sim} x_j^2 \left(\frac{1}{2}-\frac{\xi^2-1}{4\ln(\xi)}\right).\label{eq:equivalent-nu-egal-0}
\end{equation}
Hence, the PSD $P_{\nu}$ (which is inversely proportional to  $x_j^2$) tends to a constant.
At high frequencies, the Hankel's functions tend to cosine functions whose amplitudes are proportional to $x_j^{-1/2}$ and thus $P_{\nu}\propto x_j^{-2}$.

Examples of  PSD given by Eq. \ref{eq:psd-anal} are plotted in Fig. \ref{fig:PSD-anal} in function of the reduced frequency $x_j=\nu/\nu_j$, with $\nu_j=2\pi c_s/r_j$.
\begin{figure}
\includegraphics[width=0.5\textwidth]{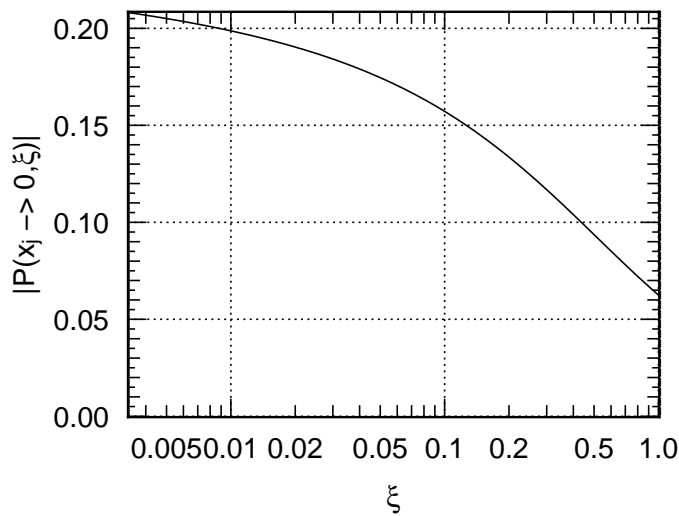}
\caption{Dependency of the reduced PSD power at lowest frequencies ($x_j \to 0$) $\displaystyle\frac{P_{x_j}(x_j \to 0)}{C_{lin}\frac{p_0}{P_0}}$ (see Eq. \ref{eq:psd-anal}) in function of the ratio $\xi=r_i/r_j$. \label{fig:pnu-a-nu-egal-0} }
\end{figure}
As expected a flat-top noise component at low frequencies and a red noise at high frequencies are present. The break frequency scales like ${ \nu_{ b }\sim 2.5 \nu_j \propto c_{\rm s}/r_{\rm j}}$ when $\xi \to 0$ and ${\nu_{ b }\sim \pi \nu_j/(1-\xi)}$ when $\xi \to 1$  (see some examples plotted in right panel of Fig. \ref{fig:PSD-anal}). To determine this value of the break frequency, we fitted the analytical PSD obtained with a zero centered Lorentzian, following the definition of \cite{belloni2002}. The PSD also exhibits a peak around the break frequency and oscillations above it as expected from Hankel's function. The ratio between the frequencies of the first peak and the second one is predicted to be close to 2. Note therefore that the frequencies depend on the value of the outer radius of the inner region. We also noticed that in our model the QPO peaks around  the break frequency.

Moreover, the PSD plateau, at low frequencies, slightly decreases when $\xi$ increases. This is due to the behavior of the modulation function ${M(x_j,\xi)}/({x_j^2(1-\xi^2)})$ at low frequency (i.e. when $x_j$ tends to 0). It is plotted versus $\xi$ for $x_j=0$ in Fig. \ref{fig:pnu-a-nu-egal-0} and shows a decrease from about 25\% for $\xi=0$ to about 5\%  for $\xi=1$.
\begin{figure*}
\centering
\includegraphics[width=0.7\textwidth]{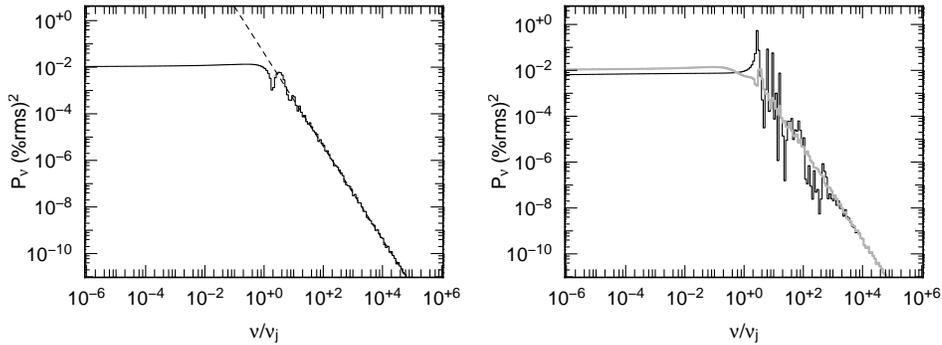}
\caption{{\it Left}: power spectrum obtained for the corona when the sound velocities in media 1 and 2 are the same ($Z=c_{s,2}/c_{s,1}=1$). {\it Right}: PSD evolution when changing the value of the acoustic impedance. Thick grey curve: $Z=1.2$, black: $Z=10$.} 
\label{fig:PSD-anal-trans}
\end{figure*}
\subsection{Wave transmission in $r_{\rm i}$}
The oscillations present above the break frequency in the PSD (see Fig. \ref{fig:PSD-anal}) are due to the infinite resonances at the eigenfrequencies of the corona due to the assumption of total reflection in $r_{\rm i}$ and no wave damping. This total reflection hypothesis can be relaxed by assuming that part of the wave is transmitted in $r_{\rm i}$ in a medium of different sound velocity. An impedance adaptation then occurs between the two media, depending on the value of the acoustic  impedance of the system $Z=c_{s,2}/c_{s,1}$, i.e. the ratio of the corona sound speed  $c_{s,1}$ to the sound speed $c_{s,2}$ below $r_{\rm i}$ . The general form of the solution in the corona (medium 1) is similar to the one obtained before (Eq. \ref{solu_J0}) and hence can be written in reduced units $x=\omega r/c_{s,1}$  as:
 \begin{equation}
p_1(x,t) = \left[\alpha H_0^1\left(x\right) + \beta
H_0^2\left(x\right)\right] e^{-\imath\omega t},
\label{eq:transm-1}
\end{equation}
whereas, in the medium 2, the transmitted wave is only progressive and hence has the following form:
  \begin{equation}
p_2(x,t) = \zeta H_0^1\left(\frac{x}{Z}\right)  e^{-\imath\omega t}.\label{eq:transm-2}
\end{equation}
In order to constrain $\alpha$, $\beta$ and $\zeta$, we need now three different equations. The first ones comes from the pressure continuity in $r_i$ and $r_j$:
\begin{eqnarray}
\alpha H_0^1(x_i)+\beta H_0^2(x_i)&=&\zeta H_0^1\left(\frac{x_i}{Z}\right){\rm,\ and}\\
\alpha H_0^1(x_j)+\beta H_0^2(x_j)&=& p_0.
\end{eqnarray}
The mass conservations in $r_i$ gives the third relation:
\begin{equation}
(\rho_1 S_1 v_1)_{r_i}= (\rho_2 S_2 v_2)_{r_i},\label{eq:transmission-conservation-masse}
\end{equation}
with $S_1=2\pi r_i h_1$ (respectively $S_2=2\pi r_i h_2$) being the vertical surface in $r_i$ in medium 1 (respectively medium 2) and $v_1$ (respectively $v_2$) the corresponding flow velocity. The link between $v_m$ and the pressure perturbations $p_m$  in each medium  is obtained by applying the Euler equations ($m=\{1,2\}$):
\begin{equation}
\rho_m \frac{\partial v_m}{\partial t}=-\vec{\nabla}p_m.
\end{equation}
Hence, by using Eqs. \ref{eq:transm-1} and \ref{eq:transm-2}, we get:
\begin{eqnarray}
-i\omega \rho_1 v_1&=&\frac{\omega}{c_{s,1}} \left[\alpha H_1^1(x)+\beta H_1^2(x)\right]{\rm \ and}\\
-i\omega \rho_2 v_2&=&\frac{\omega}{Z c_{s,1}} \left[\zeta H_1^1\left(\frac{x}{Z}\right)\right].
\end{eqnarray}
Combining these two equations with Eq. \ref{eq:transmission-conservation-masse} give then:
\begin{equation}
\frac{S_1}{c_{s,1}}\left[\alpha H_1^1(x_i)+\beta H_1^2(x_i)\right]=\frac{S_2}{Z c_{s,1}}\left[\zeta H_1^1\left(\frac{x_i}{Z}\right)\right].\label{eq:transmission-conservation-masse-2}
\end{equation} 
We need then to constrain the corona height in both side of $r_i$. We therefore assume that the corona is in hydrostatic equilibrium and in Keplerian motion. Consequently, $h/r=c_{s}/v_{Kepl}(r)$ in each medium. As a result, at the internal radius $r_i$, $S_m/c_{s,m} =
2\pi r_i h_m/c_{s,m}=2\pi r_i^2/v_{Kepl}(r_i)$. This latter value is independent from the value of $m$  and hence we obtain that $S_1/c_{s,1}=S_2/c_{s,2}=S_2/(Z c_{s,1})$. The mass conservation (Eq. \ref{eq:transmission-conservation-masse-2}) therefore reduces to:
  \begin{equation}
\alpha H_1^1(x_i)+\beta H_1^2(x_i)=\zeta H_1^1\left(\frac{x_i}{Z}\right),
\end{equation}
The resolution of the previous system gives then the full solution in pressure within the corona:
\begin{equation}
p_1(x,t)=p_0\frac{AH_0^1(x)+BH_0^2(x)}{AH_0^1(x_j)+BH_0^2(x_j)}e^{-\imath\omega
t},
\end{equation}
with,
\begin{eqnarray}
A&=&H_1^2(\xi x_j)H_0^1\left(\frac{\xi x_j}{Z}\right)-H_0^2(\xi x_j)H_1^1\left(\frac{\xi x_j}{Z}\right){\rm\ and,}\label{eq:norm-transm1}\\
B&=&H_1^1\left(\frac{\xi x_j}{Z}\right)H_0^1(\xi x_j)-H_0^1\left(\frac{\xi x_j}{Z}\right)H_1^1(\xi x_j).\label{eq:norm-transm2}
\end{eqnarray}
In the linear approximation, we can obtain the outing PSD by following the same steps as in section \ref{sec:PSD-anal}. The power spectrum from the main corona is therefore very similar to Eq. \ref{eq:psd-anal}:
\begin{equation}
P_{x_j,1} = \left|2 C_{lin}\frac{p_0}{P_0}\frac{M_1(x_j,\xi)}{x_j^2(1-\xi^2)}\right|^2,
\end{equation}
with
 \begin{eqnarray}
M_1(x_j,\xi) &=& x_j \frac{C + D}{AH_0^1(x_j) + BH_0^2(x_j)}{\rm,}\\
C &=& A\left[ H_1^1(x_j) - \xi  H_1^1(\xi x_j)\right]{\rm\ and}\\
D &=&  B\left[ H_1^2(x_j) - \xi  H_1^2(\xi x_j)\right],
\end{eqnarray}
$A$ and $B$ being given by Eqs. \ref{eq:norm-transm1} and \ref{eq:norm-transm2}.\\
 \indent Some examples of PSD are plotted in Fig. \ref{fig:PSD-anal-trans} for different values of the acoustic impedance $Z=c_{s,2}/c_{s,1}$. As expected the resonances are cancelled when there is impedance adaptation (i.e., $Z=1$) between both media (see left panel of Fig. \ref{fig:PSD-anal-trans}). The sound wave is in that case totally transmitted in $r_i$. Then the strength of the resonances are tuned by the impedance value as soon as it is different from unity and the higher the difference between both sound velocity, the stronger the resonances. Two PSD examples with $Z=1.2$ and $Z=10$ are displayed on the right panel of Fig. \ref{fig:PSD-anal-trans}). 
 \begin{figure*}
\includegraphics[clip=true,width=1.0\textwidth]{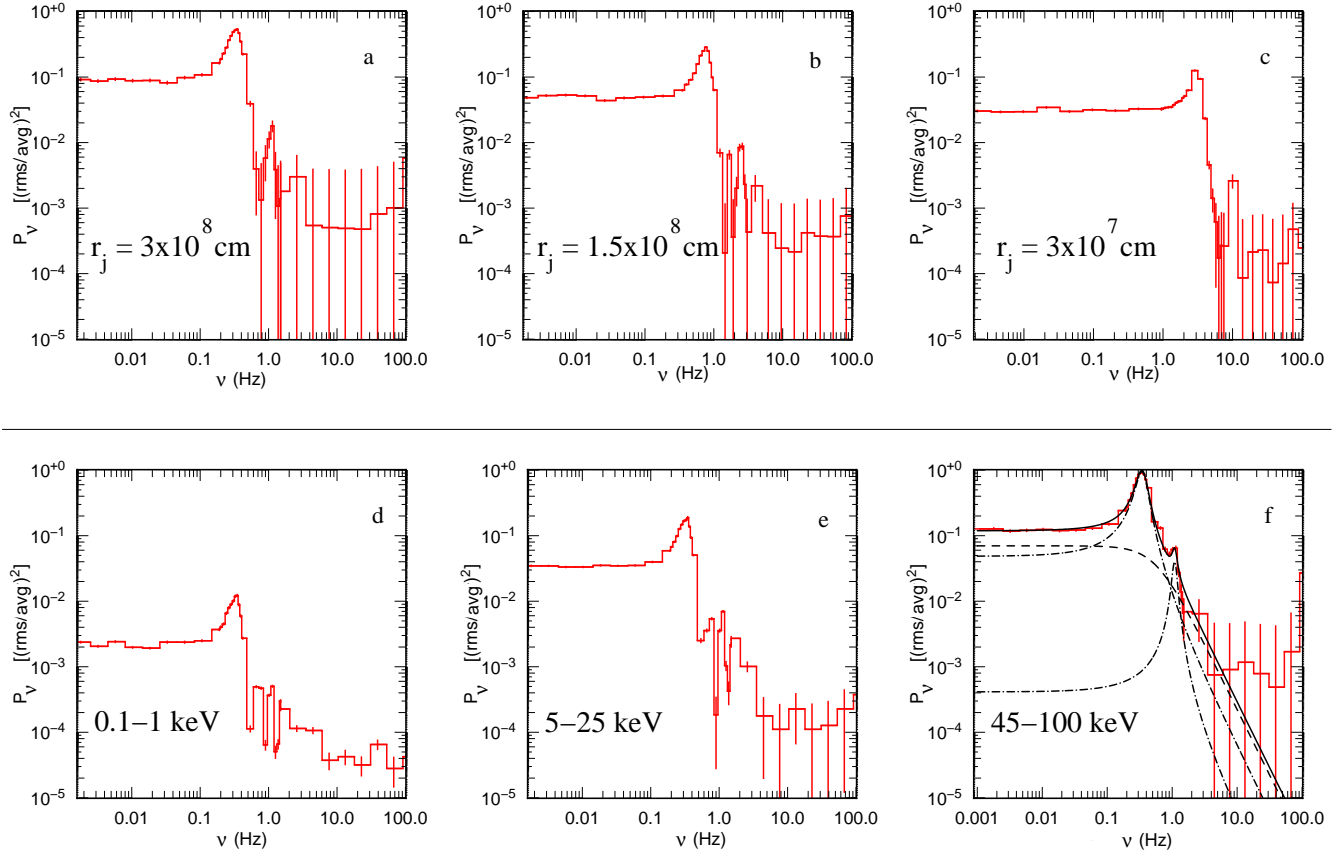}
\caption{
Evolution of the power spectrum obtained  with either $r_j$ or the energy band. $1\ \sigma$ errors arising from the fit on the amplitude of the wave in flux are also plotted. See text for the input parameters value. {\it Upper panel, from left to right:} PSD obtained for 25-45 keV band. The QPO frequency  moves upward from 0.33 to 0.67 and 3.3 Hz when the external radius of the corona $r_j$ moves inward. {\it Lower panel:} Evolution of the PSD with the energy band. The value of the external radius is $r_j=3\times 10^8\ \rm cm$. Note the increase of the overall variability with the energy. In Fig. \ref{fig:psd-1ere-simu-evolution-rayon} f The dashed curves are Lorentzians usually employed to fit the different components in PSDs of XRB.
}\label{fig:psd-1ere-simu-evolution-rayon}
\end{figure*}
\section{Monte-Carlo simulations}\label{sec:monte-carlo}
\subsection{A linear comptonisation code}
In order to check the validity domain of the linear approximation that is used in the previous section, we performed Monte-Carlo simulations of the radiative response of the corona. The comptonisation code used is linear, i.e, no feedback of the computed high energy flux on the state of the corona is taken into account. The code uses the weighted Monte-Carlo technique (see e.g. \citealp{pozdniakov1983}).  A monothermal blackbody distribution of temperature $T_{seed}$ is assumed for the seed photons which are randomly drawn in the corona mid plane ($z=0$). The temperature and optical depth of the corona at rest are fixed to $T_0$ and $\tau_0$ respectively implying a uniform pressure $P_0$. 
 The wave propagation being supposed adiabatic, the perturbation in optical depth $\delta \tau$ and temperature $\delta T$ are given by:
\begin{eqnarray}
\frac{\delta T}{T_0}&=&\frac{\gamma -1}{\gamma} \frac{p}{P_0},\ \rm{and}\label{eq:deltat-sur-t}\\
\frac{\delta \tau}{\tau_0}&=&\frac{1}{\gamma} \frac{p}{P_0}\label{eq:deltan-sur-n}.
\end{eqnarray}

Then we impose sine perturbations in $r=r_j$ with different frequencies but same relative amplitudes $\epsilon\equiv p_0/P_0$ (white noise). We assume total reflection at the internal radius and the pressure (and hence temperature and density) profile that the photon encounters while travelling within the corona is then given by Eqs. \ref{solu_J0}, \ref{eq:alpha} and \ref{eq:beta}. For those simulations, both the linear and zero-phase approximation studied in the previous section were released. Hence the photon will encounter during its travel, and scatterings after scatterings, either some part of the corona where the perturbation in pressure is positive, or negative. The number of positive and negative zones in the corona is linked to the wavelength and hence the exciting frequency. The full Klein-Nishina cross section is taken into account. The temporal evolution of the corona during the photon motion is also fully taken into account but appears to be negligible compared to the wave frequencies used, in agreement with our zero-phase approximation adopted in Sect. \ref{sec:linear}. 
We finally fit the  corona X-ray emission with a sine function in order to obtain its amplitude $L_{\sim}$ and potential phase delay. We repeat this procedure for different frequencies in order to build a power spectrum. 
\subsection{Results}\label{results}
Some examples of power spectra are plotted in Fig. \ref{fig:psd-1ere-simu-evolution-rayon} for different values of the external radius $r_j$ and different energy band indicated on the different figures. The other parameters are $r_i=3\times 10^6\ \rm cm$, $kT_0=100 \rm \ keV$, $\tau_0=1.4$ and $kT_{\rm seed}=0.25\rm \  keV$. The amplitude of the modulation in pressure is set to an arbitrary value $\epsilon=0.15$.\\

For such low value of $\epsilon$, the PSD behaviour appears in good agreement with the linear hypothesis and zero-phase approximation studied in the previous section. 
  Note that, thanks to the Monte-Carlo simulations, we are also able to study the evolution of the power spectra with energy. Interestingly, the overall shape of the power spectra can also be mimicked by using sets of Lorentzians especially at high energy as shown in Fig. \ref{fig:psd-1ere-simu-evolution-rayon} f.
\begin{figure*}
\includegraphics[clip=true,width=1.0\textwidth]{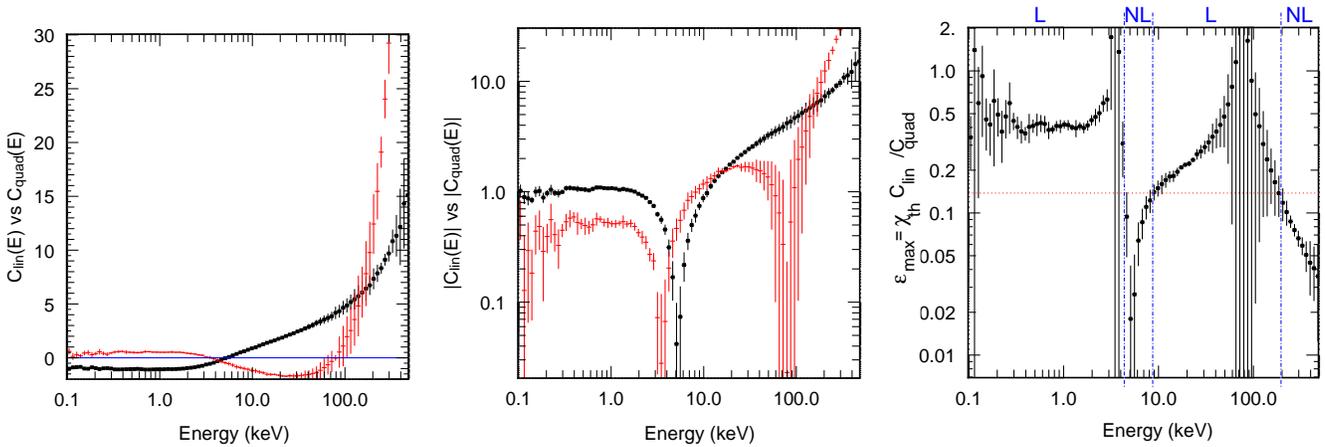}
\caption{
   \emph{Left}: $C_{lin}$ (circle) vs $C_{quad}$ (cross) value with linear coordinates on the y axis (see text for the parameter of the corona in the steady state). Under about 3 keV (the ``pivot'') the value of $C_{lin}$ is negative
This is a direct consequence of photon number conservation in the compton scattering (see text). \emph{Center}: same as right plot, but in absolute value and y axis in log coordinates, to compare the magnitude. Above 10 keV, $C_{lin}$ is higher than 1, emphasising the effect of modulation in the flux by the perturbation in pressure. \emph{Right}:  Plot of $\epsilon_{max}$ spectrum (see text for explanation). Here, $\chi_{th}=0.2$. If e.g. $\epsilon=0.15$ as plotted (dashed horizontal line), two linear and non linear zones can be observed.
}
\label{fig:Clin-vs-Cquad}
\end{figure*}
\subsection{When does the linear approximation become invalid?}
\begin{figure*}
\includegraphics[clip=true,width=1.0\textwidth]{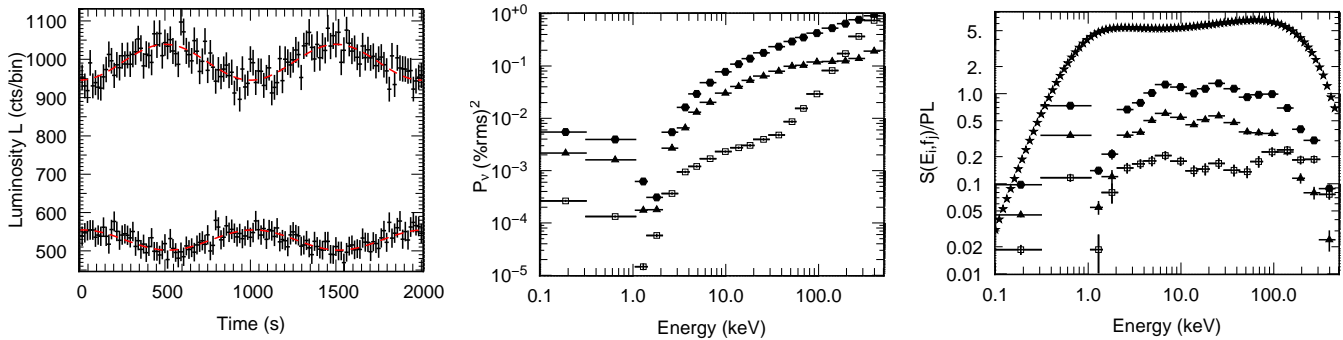}
\caption{
   \emph{Left}: Light curves obtained for the lowest frequencies probed ($\nu=10^{-3}\ \rm Hz$), for two energy bands, and their best fit by a sine function (dashed curve). Upper plot: $0.25-1 \ \rm keV$. Lower plot: $2-3 \ \rm keV$. Note the phase lag of $\pi$ between the high and low energies. \emph{Center}: ``RMS spectra'' for different frequency bands. $\blacktriangle$: $10^{-3}-10^{-1}\ \rm Hz$, $\bullet$: $10^{-1}-0.7\ \rm Hz$. {\tiny$\square$}: $0.7-1.5\ \rm Hz$. \emph{Right}: 3 lower curves: Ratio of the f-spectra to a power law model with photon index $\alpha=1.54$, for different frequencies (same symbol code as center panel). Upper curve ($\star$): ratio of the time averaged spectrum to a power law of index $\Gamma_{avg}=2.1$. Time averaged and $0.7-1.5\ \rm Hz$ spectra were rescaled for clarity.
}
\label{fig:lc-sprms-fspec}
\end{figure*}

As shown in the previous section, for low values of the modulation amplitude $\epsilon$, the Compton emission of the corona agrees relatively well with the linear approximation. We expect however some deviation from linearity for larger modulation amplitudes, deviation that should also depends on the energy. This aspect is investigated here by increasing $\epsilon$ in the simulations and then adjusting the corona spectral emission at different energy $E$ with a quadratic polynomial, i.e.:
%
\begin{eqnarray}
\left .\frac{L(\epsilon,t)-L_0}{L_0}\right |_E&\equiv& \left .\frac{\delta L(\epsilon,t)}{L_0}\right |_E\nonumber\\
&=&\left [C_{lin}(E)\epsilon+C_{quad}(E)\epsilon^2\right ]e^{-iwt} \label{polynome},
\end{eqnarray}
Hence, the energy-dependent parameters $C_{lin}$ vs $C_{quad}$ allows to ``quantify'' the linear and quadratic behaviour of the perturbation\footnote{ Note that the parameter $C_{lin}$ of Eq. \ref{polynome} is not exactly the same as the one used in Sect. \ref{sec:linear}. It also includes all the geometry-dependent part of the corona response (i.e. the term $\displaystyle\frac{2M_{x_j,\xi}}{x_j^2(1-\xi^2)}$ present in Eq. \ref{eq:calcul-analytique-composante-module-final1}) }. \\ 
\indent Starting from a reference spectrum corresponding to  $kT_{seed}=0.75\ \rm keV$, $kT_0=75\ \rm keV$ and $\tau=1.2$, we simulate 9 different spectra with values of $\epsilon$ ranging from 0 to 60\% and fit the data with the polynomial given by Eq. \ref{polynome} for different energy bins. 
The corresponding values of $C_{lin}$ and $C_{quad}$ are plotted in Fig. \ref{fig:Clin-vs-Cquad} in function of the energy. They appear strongly energy dependent. Noticeably they both cancel and change sign at a medium energy of about $2\ \rm keV$, signature of a  ``pivot'' in the variable spectral emission of the corona. \\
\indent This pivot is also clearly visible when comparing light curves at low and high energies such as those plotted in Fig. \ref{fig:lc-sprms-fspec}. Whereas the high energy lightcurve respond to the excitation coherently, the low energy one (under the pivot) has a phase lag of $\pi$. 
This pivot is due to the compton up-scattering of low energy photons which naturally produces a decrease of the number of soft photons and, simultaneously, an increase of the number of the high energy ones. Consequently  there is a $\pi$ phase lag between the two energy domains . 
The quadratic term  $C_{quad}$ cancels a second time at higher energy ($\sim 90\ \rm keV$ in the simulation plotted in Fig. \ref{fig:Clin-vs-Cquad}).\\
 
\indent By examining the trend at high energy in center panel of Fig. \ref{fig:Clin-vs-Cquad}, it is worth noting that the value of $C_{lin}$ is higher than 1 above 10 keV. This means that any fluctuation in pressure will produce an amplified radiative response of the corona at high energy.   This is due to the pivoting of the spectrum, which acts as a "lever" arm: the higher the energy bin, with respect to the pivot energy, the larger the luminosity  variation.\\  

\indent To further investigate the domains of non linearities, let's assume an arbitrary threshold $\chi_{th}$  of the ratio of the quadratic to the linear terms in Eq. \ref{polynome}, i.e. $\displaystyle \frac{C_{quad}\epsilon}{C_{lin}}$, above which we estimate that the linearity hypothesis is no more valid. This, in turn, determines different area in the $\epsilon/\chi_{th}$ vs energy plane (see Fig. \ref{fig:Clin-vs-Cquad} ) that agrees or disagrees with the linear approximation.
For example, assuming a threshold $\chi_{th}$ of 20\% means that we estimate that linearity is well satisfied as soon as  $\displaystyle \frac{C_{quad}\epsilon}{C_{lin}}<0.2$ . For the set of parameters used in Sect. \ref{results} and for which we obtain Fig. \ref{fig:Clin-vs-Cquad} this constrains the perturbation relative amplitude $\epsilon$ to be smaller than $\epsilon_{max}\simeq 50\% = 2.5\chi_{th}$ in the $0.1-3$ keV energy band and even much smaller above 10 keV. For high values of $\epsilon$, say e.g. $\epsilon=0.7$, the whole energy domain is non-linear. For lower values, say e.g. $\epsilon=0.15$ as plotted by the horizontal line in right Fig. \ref{fig:Clin-vs-Cquad}, two linear and two non linear zones can be noted. 
\subsection{Energy dependence of the variability}
Finally, the strength of the variability and its evolution with energy is investigated. For that purpose, we computed and plotted on Fig. \ref{fig:lc-sprms-fspec} the so called ``RMS spectra'' and ``f-spectra'' . Those ``variability spectra'' were computed for three spectral ranges: at the lowest frequencies, where the PDS is flat ($10^{-3}-10^{-1}\ \rm Hz$), near the resonance ($10^{-1}-0.7\ \rm Hz$) and after ($0.7-1.5\ \rm Hz$).\\
\indent As expected from the behaviour of $C_{lin}$, the presence of the pivot lead to a drastic decrease of the rms at around 2 keV for all the frequencies studied, which is a strong prediction of the model, even if the position of this pivot in energy depends on several input parameter such as the seed photon temperature. After it, the variability increases without any cut-off at high energy. This due to the fact that the wave in the corona also modulates the temperature of the high energy electrons and hence the position of the cut-off in the high energy spectrum. The higher the modulation, the larger the cut-off. An increase of the variability of the $0.7-1.5\ \rm Hz$ component is also predicted above $\sim 100\ \rm keV$.\\
\indent F-spectra were also computed, following definition of \cite{revnivtsev1999}, and adopting the same three frequency bands as above. Such spectra were then fitted between 2 and 150 keV by a powerlaw. Their photon indices are decreasing with increasing frequencies as they are equal to $1.61\pm0.02$ for the $10^{-3}-10^{-1}\ \rm Hz$ band, $1.53\pm0.01$ for the $10^{-1}-0.7\ \rm Hz$ band and  $1.48\pm0.05$ for the $0.7-1.5\ \rm Hz$ band. In contrast, the time averaged spectrum has a higher photon index ($\Gamma_{avg}\sim 2.1$). The ratio of the obtained f-spectra to a powerlaw of index $1.53$  is plotted in Fig. \ref{fig:lc-sprms-fspec}, and also the ratio of the time averaged spectrum to a powerlaw of index $2.1$. The time averaged and the f-spectra exhibit a quite sensible deviation from the simple power-law model. This discrepancy with a power-law is however more emphasised in the f-spectra than in the time averaged spectra.
\section{Discussion}\label{sec:discussion}
The timing response of an oscillating corona in cylindrical geometry is investigated, the corona acting as a filter when responding to a perturbation in pressure generated at its external radius. In the case of total reflection at the internal radius, infinite resonances may occur. They may be damped in the case of acoustic impedance adaptation between the corona and the external medium.
We have shown that the filtering effect of the corona leads to power spectra with a broad band shape well fitted by a Lorentzian centered in $\nu=0$ (see Fig. \ref{fig:psd-1ere-simu-evolution-rayon}). Moreover, the resonances produce a major peak at the PSD break and smaller peaks at higher frequency. The characteristic frequencies of the PSD break and peaks scale with $\nu_j=2\pi c_s/r_j$. For a 10 solar mass black hole, an outer corona radius $r_j=200\rm\ R_g$ and a corona temperature $kT_0=100\ \rm keV$ (i.e. a sonic velocity $c_s=3.1\times 10^8\ \rm cm.s^{-1}$), this gives $\nu_j=0.17\ \rm Hz$.
These  power spectra are thus relatively similar to the one observed in the HIMS (Hard Intermediate State) of X-ray transients, whose shape are well fitted by 3-4 Lorentzians peaking at different frequencies from a tenth of hertz (for the break and the low frequency QPO, hereafter LFQPO) to hundreds of hertz (for the high frequency QPO, see e.g. \citealp{nowak2000} or \citealp{pottschmidt2003}). \\
Interestingly our low frequency peak  is always of the order of the frequency break $\nu_b$ (both scaling with $\nu_j$). This is also in agreement with the observed correlations between $\nu_b$ and the low frequency QPO $\nu_{LFQPO}$ \citep{wijnands1999a,belloni2002,kleinwolt2008}. We note however that $\nu_{LFQPO}$ is usually 5 times higher than $\nu_b$, especially in \cite{wijnands1999a} relationship. We note however that in the equivalent correlation plotted on Fig. 11 of \cite{belloni2002}, the lower branch shows a correlation between the break frequency $\nu_b$ and the typical frequency of the ``hump'' ($\nu_h$), with a ratio close to one. Finally, the ratio between the first two peak frequencies is close to 2 in our model, as it is observed for several type C QPO.\\ 

\indent Our model also predicts a decreasing PSD power and an increasing resonance peak frequencies when decreasing the outer corona radius $r_j$. This is here again consistent with the general trend observed in BHB where the PSD power decreases when the observed frequencies increase (BH anti-correlation, \citealp{belloni1990}). However in our model, this effect is low and could hence only account for a fraction of the observed anti-correlation. Indeed, the BH anti-correlation translates the fact that in $\nu P_{\nu}$, the different observed Lorentzians are peaking at roughly the same level, or equivalently that the rms integrated variability remains roughly constant. An extra condition on the exciting process is therefore required, with a rms level that must decrease when the size of the corona decrease as well. The necessary condition would hence be that  $\delta P\propto r_{\rm j}^{\alpha}$, with $\alpha\geq 0.5$.\\ 

 \indent As shown in Fig. \ref{fig:lc-sprms-fspec} we expect a $\pi$ phase lag between the very low energy and the high energy light curves, which is a strong prediction of the model. This is a direct consequence of the assumed geometry, especially of the fact that the source of seed soft photons is assumed at the corona midplane. This source is necessarily on the line of sight of the observer. Then due to the photon number conservation during the Compton scattering process, the disappearance of the soft photons is directly compensated  by an increase of the hard ones. Moreover it depends on the pivot energy which crucially depends on the input parameters (especially $kT_{c}$ and $kT_{soft}$).  In a geometry where the soft seed photons are produced outside the corona, this constraint can be easily relaxed. This would be the case of a soft photon field produced by an outer accretion disc. Such geometry (inner hot corona surrounding by an outer accretion disc) is indeed generally believed to qualitatively well reproduced the inner region of the accretion flow around compact objects (i.e. \citealt{esin1997,done2007}). Note that with such geometry, a decrease of the outer corona radius $r_j$ (which corresponds to the inner accretion disc one) would imply an increase of the seed photons temperature and flux i.e. an increase of the corona coolings. Consequently a softening of the X-ray spectrum should happen. Since a decrease of $r_j$ implies also, in our model, an increase of the resonance peak frequencies, a correlation between the X-ray photon index and the QPO frequencies is expected. Such correlation is indeed observed in different XRB \citep{titarchuk2005}. \\
\indent The position of the pivot can also be drastically changed and moved towards lower energies if the seed photons typical energy is far lower than considered here. For example, as suggested e.g. in \cite{malzac2009}, the soft photon contribution could originate from the synchrotron emission in optical and UV from the hot plasma.

\indent We also demonstrated that $C_{lin}$ increases after the pivot whereas $C_{quad}$ is decreasing again around $20\ \rm keV$. The general increase of $\delta L/L$ with the energy after the pivot (visible e.g. in the rms-spectra plotted in Fig. \ref{fig:lc-sprms-fspec}) could be responsible in part for the observed increase of the variability with energy for the QPO and the continuum (see e.g. for XTE J1550-564 \citealp{cui1999b}). Alternatively, the observed decrease of the QPO harmonic strength in Fig. 5 of \cite{cui1999b} could also be well explained by the drop in the value of $C_{quad}$ after $30\ \rm keV$. We can also note the similarity between the RMS spectra plotted in Fig. 5 of \cite{cui1999b}, when the source is in its HIMS, and the one we obtained with our model. The increase of the QPO and continuum variability between 2 and 20 keV is then a direct consequence of the comptonisation process.\\
\indent We also observe a decrease of the powerlaw slope in the f-spectra when the frequencies get higher. This is consistent with the results obtained for Cyg X-1 in \cite{revnivtsev1999}. It is worth noting that those authors explain the presence of ``wiggles'' in the f-spectra as a consequence of reflexion features on an optically thick material. Our Monte-Carlo simulations shows however that those features can be generated by the comptonisation process only.\\
\indent The non-linear domain is then investigated as a function of the energy. We demonstrated that if a reasonable arbitrary value of the non-linear to linear ratio is chosen, it is expected that the non-linear effects of the radiative response will occur only at very high and medium energies.\\
\\
\indent The choice of a cylindrical geometry has also an impact on the degree of loss of the input signal due to scatterings. The following arguments are similar to those discussed in previous studies, such as in e.g \cite{miller1995}.
Indeed, unless one global oscillation takes place in the corona (which is the case at very low exciting frequencies), a photon may encounter alternatively positive and negative perturbations in pressure if it travels in the radial direction. It would consequently kill the effect of the input signal on the output lightcurve. We even expect this effect to be emphasised when looking at high energies, as, regardless of the optical depth of the medium chosen, a high energy photon encountered a large number of scatterings. On the contrary, the oscillation will not be smeared out if the travel of the photon remains local.\\
\indent As we chose a cylindrical geometry, the optical depth $\tau\sim 1$ is relative to the vertical direction and hence this implies that in the azimuthal and radial directions $\tau_{r,\theta} \gg 1$. As a consequence, this is forcing the locality, relative to the radial wavelength of the pertubation, which in turn preserves the signal, whatever the output photon energy. As one goes towards a more spherical geometry, where $\tau$ now becomes a measure of the radial optical depth, $\tau$ in various directions
becomes much less anisotropic.  Looking at higher energies means
looking at a larger range of sampled locations, and hence smearing of
the imparted variability signal for sufficiently small wavelengths of
the perturbation. 
\\
\\
\indent While this simple model already provides very interesting timing behavior and appears very promising to explain the main features observed in the PSD of XRB, significant improvements are needed to include different, and potentially important, physical effects. \\
\indent For instance we have shown that the effect of transmitting the wave in an inner medium results in softening the resonances and the gaps in the power spectra. But the kinematics of the corona could also play an important role. Assuming the corona located in the inner region of an accretion flow, some (differential) rotation is expected and we might expect the emitted frequency by each part of the corona to be Doppler shifted, smoothing the PSD in the same way emission lines can be smoothed in energy spectra. In the close vicinity of the central engine, the general relativistic effects should also significantly contribute to blurred the PSD. The use of a non uniform sound velocity profile (due to e.g. non uniform corona temperature) along the radial or vertical directions is also expected to have some impact in this respect.
But more importantly, including all these effects will result in a very different dispersion equation followed by the acoustic waves and thus to a significantly different behavior of the radiative response of the corona. \\
\indent We also used a single sound wave, but other acoustic waves could be present. For example, in the case of magnetic plasma Alfv\'en and both slow and fast magnetosonic waves, with velocities $v_A$, $c^-$ and $c^+$, should be used. The former do not generate density or pressure perturbation. They are not expected to change the comptonisation efficiency, contrary to the magnetosonic ones. In the case of plasma in equipartition, $c_s^2=v_A^2$, hence $c^-=0.54 c_s$ and $c^+=1.31 c_s$ and we do not expect significant changes compared to the sound wave case studied here\footnote{The above calculations were done with $v_{A,\ poloidal}=1/\sqrt{2}\ v_{A}$.}. However, for plasma far from equipartition,  $c^- \sim c_s\ne c^+\sim v_A$. Therefore if both wave propagates in the medium, each component would be responsible for its own band limited noise and resonances in the power spectra. Multiple perturbating waves may then explain the multiple Lorentzian components generally needed to fit  the PSD  of XRB in hard states (e.g. \citealt{nowak2000,pottschmidt2003}). Note however that in the HIMS, only two Lorentzians are necessary. If we identify in our framework the low frequency Lorentzian ($L_{b}$, see \cite{belloni2002} for the labels of the different PSD components) with the propagation of a slow magnetosonic wave and the lower upper frequency ($L_l$) with the fast one, it would require a very high magnetisation parameter for the plasma. Indeed the observations gives roughly $\nu_l\sim 50 \nu_b$ (see e.g. tables 2 and 3 in \citealp{belloni2002}), and since $c^- \sim c_s$ and $c^+\sim v_A$, it would imply that $v_A^2/c_s^2 \sim 2500$.\\

\indent We also did not investigate the implied time-lags generated by the comptonisation process. It is however known that such lags can be reproduced by pivoting in the spectra (see e.g. \citealp{poutanen1999} or \citealp{koerding04}), and our models predicts the presence of such a pivot.\\
\indent More fundamentally, as we probe the timing response of the corona, it can be used as a complement to other variability models as the shape of the input excitation is supposed here to be a white noise. The RMS-flux scaling found in both AGN and XRB (see e.g. \citealp{uttley2001}) has demonstrated that the processes at the origin of the variability is, by essence, non-linear \citep{uttley2005}. This is not necessarily in contradiction with our study since we only investigate what is the timing response of the corona, without taking into account the feedback process of the corona on the disc itself. Our model just examine the subsequent effects of the corona and in particular how the radiative transfer and the geometry of this optically thin medium could filter any input variability.\\
\indent The present model does not intend to reproduce all the observed features concerning XRB variability, especially due to its simpleness. The results obtained in this paper are however very promising for further investigations, and a more detailed analysis, including the effect of the corona rotation and geometry will be presented in a forthcoming paper.
\section*{Acknowledgements}
C.C thanks J. Rodriguez, D. Hannikainen, D. Barret, A. Goldwurm, J.-M. Hameury, C. Done, P. Uttley, I. Papadakis and R. Belmont for useful discussions on the subject. The authors also thank the anonymous referee for his useful comments on the paper. This work has been supported by the French National Agency (ANR) through the project ``Astro2flots'' ANR-05-JCJC-0020. T.B. acknowledges support from contract PRIN INAF 2008.
\appendix
\renewcommand{\thefigure}{\Alph{figure}}
 \setcounter{figure}{0}
\renewcommand{\thetable}{\Alph{table}}
 \setcounter{figure}{0}
\section{Low frequency dependence of the modulating function $M$}\label{app:A}
Hankel's functions can be expressed in function of Bessel's one:
\begin{eqnarray}
H_n^1(x)&=&J_0(x)-i Y_0(x)\\
H_n^2(x)&=&J_0(x)+i Y_0(x)
\end{eqnarray}
For low values of $x$, the equivalents of the Bessel's function are the following: 
\begin{eqnarray}
J_0(x)&\sim& 1\label{eq:equiv-j0}\\
 Y_0(x)&\sim& \frac{2}{\pi} \left(\ln\left(\frac{x}{2}\right)+\gamma\right)J_0(x)\label{eq:equiv-y0}\\
 J_1(x)&\sim& \frac{x}{2}\label{eq:equiv-j1}\\
Y_1(x)&\sim& - \frac{2}{\pi x} + \frac{x}{\pi}\ln\left(\frac{x}{2}\right) -\frac{x}{2\pi}(-2\gamma+1)\label{eq:equiv-y1}
\end{eqnarray}
 \citep{abramowitz1964}, where $\gamma$ is the Euler-Mascheroni constant. If the Hankel's functions are expressed in terms of Bessel's functions, both numerator and denominator of the function $M$ (Eq. \ref{eq:calcul-analytique-composante-module-final2}) are pure imaginary and the modulating function $M$ writes:
\begin{eqnarray}
M(x_{\rm j},\xi)&=&x_{j}\frac{J_{0,\rm i}(Y_{1,\rm j}-\xi Y_{1,\rm i})-Y_{0,\rm i}(J_{1,\rm j}-\xi J_{1,\rm i})}{J_{0,\rm i}Y_{0,\rm j}-Y_{0,\rm i}J_{0,\rm j}}\\
&=&x_j\frac{N}{D}.\label{eq:B-N-D}
\end{eqnarray}
Using the four equivalents \ref{eq:equiv-j0}, \ref{eq:equiv-y0}, \ref{eq:equiv-j1} and \ref{eq:equiv-y1}, the numerator $N$ and denominator $D$ writes:
\begin{eqnarray}
N&=&\frac{-2}{\pi x_j}+\frac{2\xi}{\pi \xi x_j}+\frac{x_j}{\pi}\ln\left(\frac{x_j}{2}\right)-\frac{\xi^2 x_j}{\pi}\ln\left(\frac{\xi x_j}{2}\right)\nonumber \\
& & -\frac{x_j}{2\pi}\left(-2\gamma+1\right)+\frac{\xi^2x_j}{2\pi}(-2\gamma+1)\nonumber\\
& & -\frac{2}{\pi}\left(\ln\left(\frac{\xi x_j}{2}\right)+\gamma\right)\left(\frac{x_j}{2}-\frac{\xi^2 x_j}{2}\right) \\
D&=&\frac{2}{\pi}\left (\ln\left (\frac{x_j}{2}\right )\right )-\frac{2}{\pi}\left (\ln\left (\frac{\xi x_j}{2}\right )\right ).
\end{eqnarray}
Those equations can be rewritten as:
\begin{eqnarray}
N&=&\frac{x_j}{\pi}\left[\ln\left(\frac{x_j}{2}\right)-\xi^2\ln\left(\frac{\xi x_j}{2}\right)\right. \nonumber \\
& &-(1-\xi^2)\frac{-2\gamma+1}{2} - \ln\left(\frac{\xi x_j}{2}\right) -\gamma \nonumber \\
& &\left . +\xi^2 \ln\left (\frac{\xi x_j}{2}\right ) +\gamma \xi^2 \right ]\\
D&=& -\frac{2}{\pi}\ln\left(\xi\right).
\end{eqnarray}
Finally, we get:
\begin{eqnarray}
N&=&\frac{x_j}{\pi}\left[-\frac{1-\xi^2}{2}-\ln(\xi) \right]\\
D&=& -\frac{2}{\pi}\ln\left(\xi\right).
\end{eqnarray}
These last equations combined with Eq. \ref{eq:B-N-D} lead to the result written in Eq. \ref{eq:equivalent-nu-egal-0}.
\bibliographystyle{mn2e}
\bibliography{ref}
\label{LastPage}
\end{document}